\documentclass[manuscript]{acmart}
\AtBeginDocument{%
  \providecommand\BibTeX{{%
    \normalfont B\kern-0.5em{\scshape i\kern-0.25em b}\kern-0.8em\TeX}}}

\setcopyright{rightsretained}
\copyrightyear{2021}
\acmYear{2021}
\acmDOI{10.1145/3397482.3450718}

\acmConference[IUI '21 Companion]{26th International Conference on Intelligent User Interfaces}{April 14--17, 2021}{College Station, TX, USA}
\acmBooktitle{26th International Conference on Intelligent User Interfaces (IUI '21 Companion), April 14--17, 2021, College Station, TX, USA}
\acmDOI{10.1145/3397482.3450718}
\acmISBN{978-1-4503-8018-8/21/04}

\begin{document}

\title{TableLab: An Interactive Table Extraction System with Adaptive Deep Learning}

\author{Nancy Xin Ru Wang} 
\affiliation{% 
\institution{IBM Research} 
\city{San Jose} 
\country{USA} 
} 
\email{wangnxr@ibm.com} 

\author{Douglas Burdick} 
\affiliation{% 
\institution{IBM Research} 
\city{San Jose} 
\country{USA} 
} 
\email{drburdic@us.ibm.com} 

\author{Yunyao Li} 
\affiliation{% 
\institution{IBM Research} 
\city{San Jose} 
\country{USA} 
} 
\email{yunyaoli@us.ibm.com} 
\renewcommand{\shortauthors}{Wang et al.}

\begin{abstract}
  Table extraction from PDF and image documents is a ubiquitous task in the real-world. Perfect extraction quality is difficult to achieve with one single out-of-box model due to (1) the wide variety of table styles, (2) the lack of training data representing this variety and (3) the inherent ambiguity and subjectivity of table definitions between end-users. Meanwhile, building customized models from scratch can be difficult due to the expensive nature of annotating table data. We attempt to solve these challenges with TableLab by providing a system where users and models seamlessly work together to quickly customize high-quality extraction models with a few labelled examples for the user's document collection, which contains pages with tables.  
%Users in TableLab are recommended a few table examples already extracted with a pre-trained base deep learning model. These examples are automatically selected as representative samples of the dataset as we detect templates from the input collection by clustering embeddings from the extraction model. In an easy-to-use and modular user interface, users provide feedback to these detections and these are applied to finetune the models. These customized models can then be applied to the document collection with improved table extractions than the base model.
Given an input document collection, TableLab first detects tables with similar structures (templates) by clustering embeddings from the extraction model. Document collections often contain tables created with a limited set of templates or similar structures. It then selects a few representative table examples already extracted with a pre-trained base deep learning model. Via an easy-to-use user interface, users provide feedback to these selections without necessarily having to identify every single error. TableLab then applies such feedback to finetune the pre-trained model and returns the results of the finetuned model back to the user. The user can choose to repeat this process iteratively until obtaining a customized model with satisfactory performance.  
\end{abstract}




\maketitle

\section{Introduction}

\subsection{Challenges TableLab Addresses}
Recently, there has been increasing interest in extracting complex structures such as tables from PDF and image documents~\cite{burdick2020table}. Table extraction involves identifying the border and the cell structure for each document table such that it can be displayed in a structured format like HTML. The motivation for TableLab came from requests from industry professionals for the ability to easily create ground truth data and customize models for extracting tables for their specific document collections.  TableLab accomplishes this by addressing the following table extraction challenges.    

\begin{figure*}[h]
        \centering
        
        \includegraphics[width=0.6\textwidth]{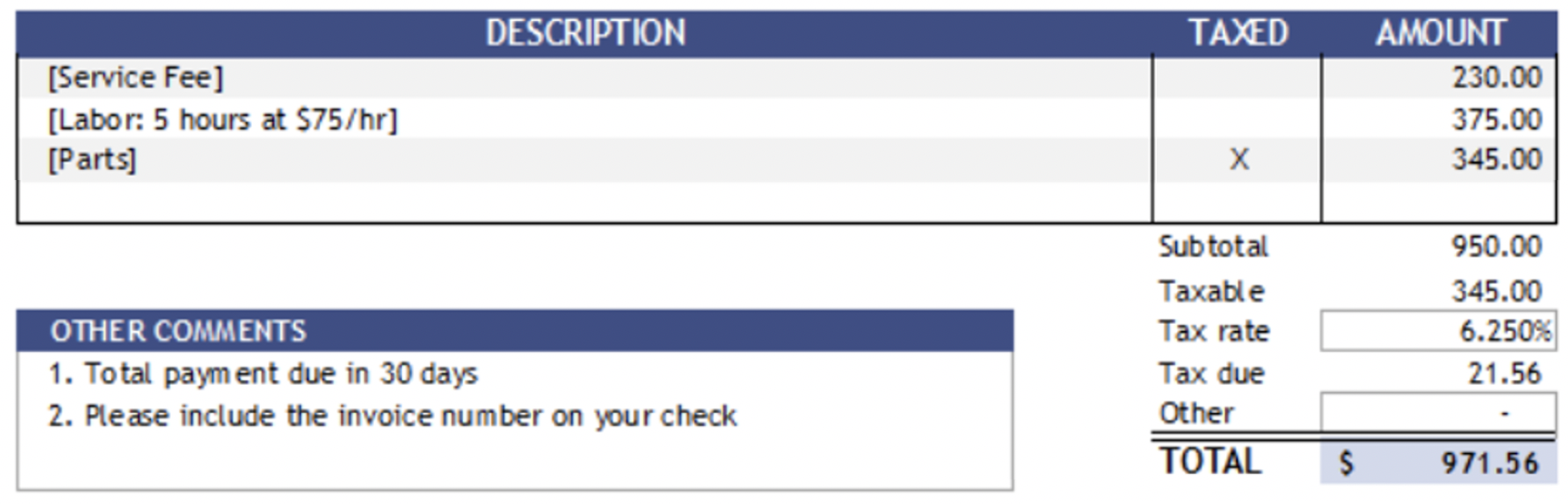}
        \caption{Example document with ambiguous tables. Whether the main invoice table should be one or two tables will depend on downstream tasks for the user.}
        \label{fig:example}
    \end{figure*}

First, there is great diversity of table formatting across different documents types and sources.  Tables from invoices are formatted differently than those from scientific articles or financial reports, with the visual clues across sources providing conflicting information about the table border and/or structure. Thus, creation of a single high-quality model to support table extraction from the wide diversity of document types is difficult if not impossible when considering the fact that even humans can disagree about table definitions from the same source document (see Figure.~\ref{fig:example}). Despite the diversity of table formats encountered in real-world settings, the user's needs and table extraction expectations are ultimately the most important. TableLab leverages this observation by supporting finetuning of a high-quality table extraction model trained on hundreds of thousands of tables using a small number of user labelled examples.

%Additionally, table-specific machine readable markup is unavailable for many widely used document formats, such as PDF or scanned images.  The most prevalent information available from such document formats is text data, including formatting (font size, color, bold/italic, etc.) and positional information (i.e., coordinates of text on page) with text values, and graphical elements (e.g., line segments). %Thus, the space of features for any table extraction algorithm is limited.  
%The extraction of such data is often error-prone (e.g., OCR for recognizing text from images, parsing programmatic PDF)
%, and requires correction before the text data can be presented to the table extraction algorithm. 
%This issue becomes more pronounced for low-quality source documents. TableLab supports correction of extracted text.
%, with correction operators including adding missing text, removing spurious text and editing incorrect text.  
%Such corrected text data could serve as ground-truth labelled data for improving text extraction approaches (e.g., train / fine-tune OCR models for text recognition).  

TableLab also supports efficient labelling of tables in documents, which involves two sub-problems.  First, how do we select the most useful examples for labelling which improve finetuned model accuracy the most?  Second, how do we effectively label individual example tables, particularly table structure where the same error repeatedly occurs?  The mechanisms TableLab uses to improve labeller efficiency are described in the overview section.      

%\begin{itemize}
    %\item Define problem
    
\subsection{Related Works}
    Many deep-learning solutions have been applied to the table extraction problem in recent years.
    Examples for table and cell region detection include~\cite{Schreiber2017DeepDeSRTDL,Li2019TableBankTB,Gilani2017TableDU} while ~\cite{Schreiber2017DeepDeSRTDL,tensmeyer2019deep, prasad2020cascadetabnet, Li2019TableBankTB, zhong2019image} address table structure extraction. 
    However, none of these solutions are able to extract tables exactly for all documents from all domains due to the wide variety and ambiguity of the problem (See Figure~\ref{fig:example} for an example)~\cite{burdick2020table, Hoffswell2019InteractiveRO}. Additionally, labelled data is tedious to create. There are a few existing large-scale datasets for scientific papers~\cite{zhong2019image} and financial reports~\cite{zheng2020global} but many documents in business are confidential. 
    %In this work, we propose TableLab, which is a system that integrates deep learning models with human feedback in order to improve extraction results and customize models for a particular user and document collection. 
    %\item Discuss related work in Table Extraction
    
    %Table extraction has been tackled by traditional rule-based, and more recently, deep learning approaches. Several deep learning approaches borrow from the object recognition community for table and row detection. There is also a variety of methods for structure recognition, ranging from position based clustering to graph neural networks. 
    %\item Discuss related work in Labelling for tables
    Research into ease of labelling and active learning for tables is not as well studied as table structure extraction. Hoffswell et al. ~\cite{Hoffswell2019InteractiveRO} design a system to help users repair extracted tables with a mobile interface. However, users are unable to directly improve the extraction model with their annotations. 
    %\item Differentiation factor:

\subsection{TableLab Design Considerations}
    Current table extraction systems extract tables without the option to give feedback. Since the systems do not work well on all document types, users can be frustrated by the lower quality of extractions without the ability to improve them. Our system finetunes models in an iterative fashion, collaborating with the model to quickly label and see improvements with the model. In our system, we first use deep learning models to extract table and cells to generate table structure. Using visual embeddings derived from the model, we cluster documents into templates in order to recommend specific pages to label for users that balances between ease of labelling and the most impact on the model with a large variety of styles. Thanks to this recommendation system, we minimize the size of labeled data required. As well, since our table extraction model is modular in nature, some labels for components (ex. table border) can immediately improve results in others (cell border) such that the user does not need to repair every error in the table extraction process. 
    %\begin{itemize}
    %    \item Use ML model to first extract result
    %    
    %    \item Use features to cluster documents to into templates to recommend ones to %label for users
    %    \item Modularize labelling process to make it easier for users
    %    \item Size of labeled data required
    %    \item Pipeline composition
    %\end{itemize}
    %\item Dataset for video
    %\begin{itemize}
    %    \item Pretrained on scientific docs (Pubtabnet)

    %\end{itemize}
%\end{itemize}

\section{Overview of TableLab}
\begin{figure*}[h]
        \centering
        \includegraphics[width=0.8\textwidth]{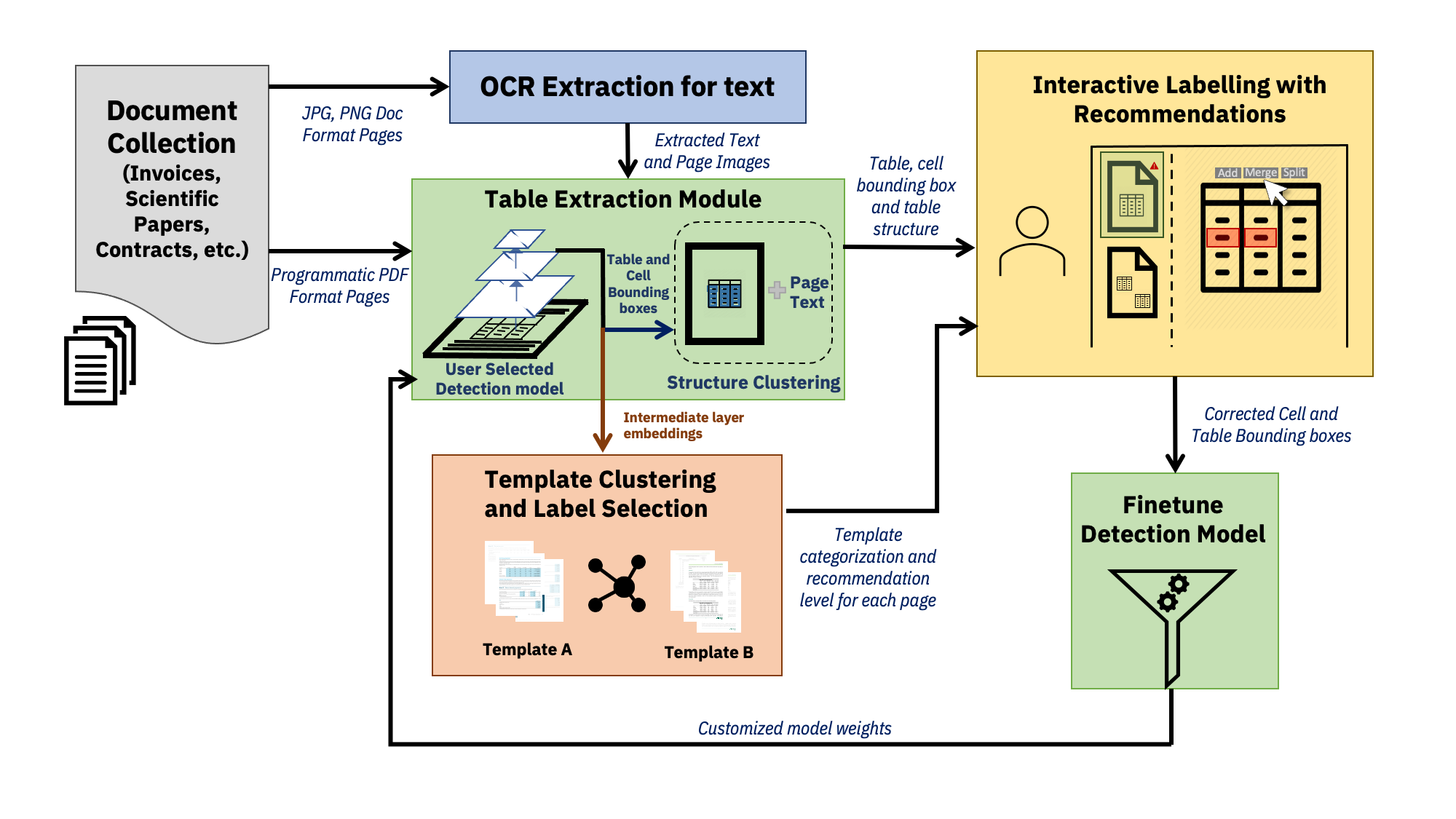}
        \caption{System architecture of TableLab. Details for each step are described in the overview section.}
        \label{fig:arch}
    \end{figure*}
%\begin{itemize}
    \subsection{Table Extraction}
    To begin, we apply our table extraction module (based on the GTE framework~\cite{zheng2020global}) to the user's document collection.
    %Our module first consists of 2 separate object detection models for tables and cells. 
    We provide a few base model weights that have been pre-trained on different document types for users to select the one that best matches their collection. After the deep models have been applied, we input the resulting table and cell bounding boxes as well as the document text snippets (scanned and image documents are first processed with an OCR engine in order to extract the text) into our structure clustering model.  This model determines the row and column assignment and the content of each cell such that it can be represented by a structured format such as HTML. 
     \subsection{Template Clustering and Label Selection}
    %\item Apply model to collection \& serve recommended pages to label
    After the deep learning models have been applied to the collection, the visual embeddings from the detection models are used to cluster the document collection into templates. After clustering, the lowest and highest confidence pages of each template is selected for user labelling. The labels for lowest confidence pages will provide the most benefit to the model while the highest confidence pages should be easy and quick to label, allowing for faster feedback. An icon for each label recommendation type is indicated beside each page in the user interface.
    %\item Easy to use UI for giving recommendations
    
    \begin{figure*}[h]
        \centering
        \includegraphics[width=0.75\textwidth]{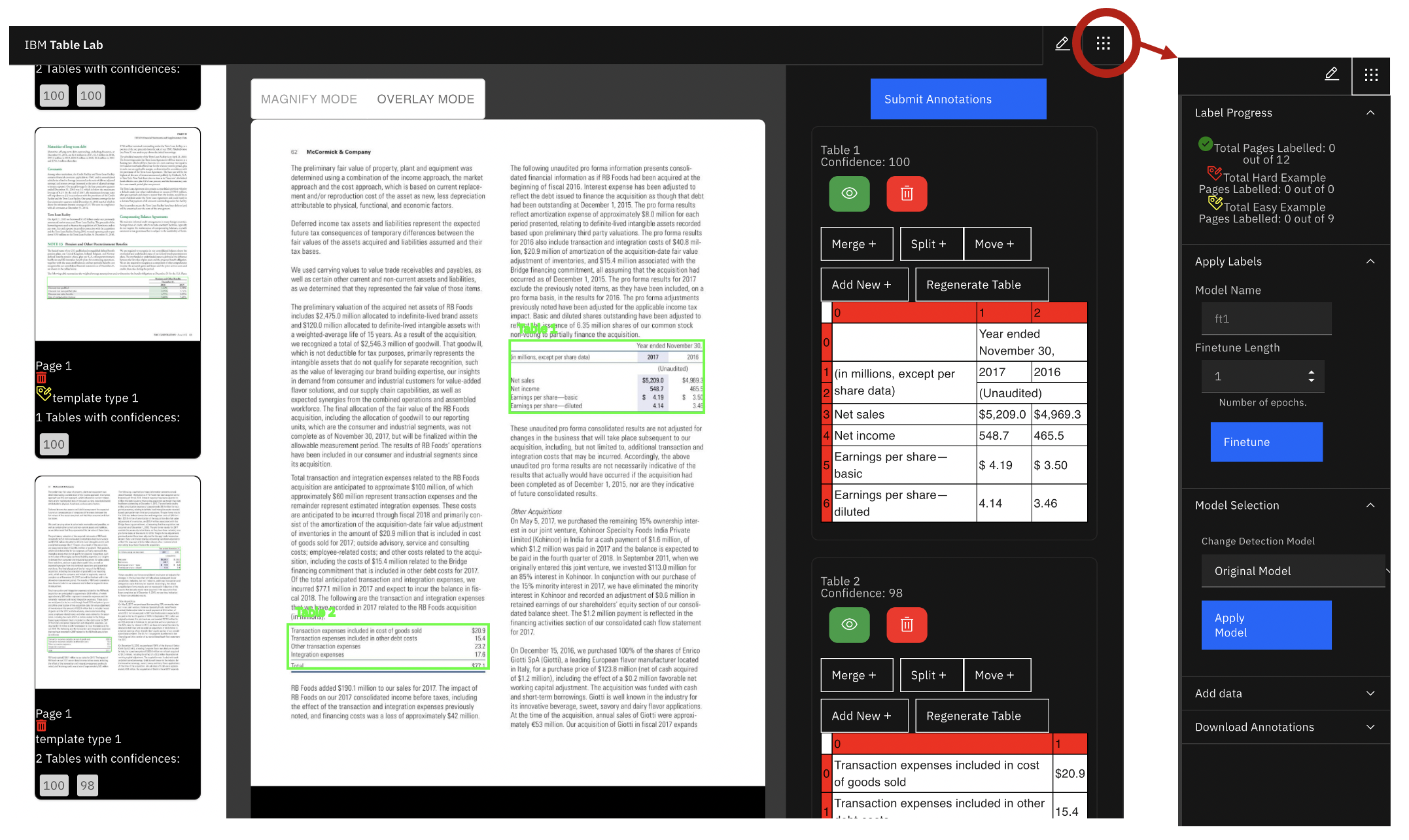}%
        \caption{TableLab in model selection and table editing view. The leftmost panel contains previews of each page in the input document collection with detected table boundaries, their confidence, template type as well as a yellow or red tag if they are recommended for labelling. In center, the user may view the page in the table box (and text when a textbox is selected) Overlay mode as shown or they may magnify the contents in Magnify mode. The user can interact with the boxes to change their size and location. The right panel shows the extracted table and our table editing features. We also show the additional panel where users can view their labelling progress, finetune models with labelled data, select the model used to extract tables, add additional documents and download the table annotations.}
        \label{fig:edit_screenshot}
    \end{figure*}
    \subsection{Interactive Labelling with Recommendations}
    When the initial table structure has been extracted and label recommendations determined, the user will be able to view the extracted tables and provide feedback as needed. In a typical case, the user can first adjust the table border. This prompts the system to redo the structure clustering, providing an updated extracted table. Sometimes, this results in a completely correct extraction and the user may submit the page for finetuning at this time. Otherwise, the user has full control to merge, split cells or whole columns and rows similar to manipulations in a spreadsheet program. By leveraging the layout of the text snippet positions, users can split and move cell content by text chunks rather than word by word. The user may also edit a text snippet by typing in the content and adjusting its bounding box. 
    \subsection{Model Finetuning}
    When the page has the correct table extraction, the user may submit the page for model finetuning and apply the customized model to their collection for improved extraction results. If there are additional errors, the user may make additional corrections and repeat the finetuning process. For a typical collection, we find that one finetuning round is generally enough to correct the rest of the collection but this depends largely on the diversity of the collection itself. 
    
    %Our system provides an easy to use UI to manipulate table boundaries, cell structure, as well as textual content. 
    %Since there are separate models for table and cell that can precompute portions of the table extraction pipeline, user feedback for some components (like table border) can be immediately integrated with cell information to regenerate new table structure. In some cases, no further feedback is required. 
    %. Additionally, the clustering algorithm of the table structure recognition can use limited feedback to improve the structure itself such that the user does not need to label the full table. Rather, it is an iterative process to improve the models with a few pieces of feedback from the user. 
    %\item Easy to finetune models and view new results in same system and model and Tannotations can be downloaded for future use
    
    %When the user has completed corrections for all the recommended pages or when the user feels they have labelled enough pages, the user can indicate a name for their custom model select the finetune model button. After training, the user can apply the new model to the remaining unlabelled pages. 
    
   % Once the user is satisfied with the quality of the extraction, the models can be downloaded and applied in a parallel cluster for faster throughput on the full document collection and for future use. The annotations can also be downloaded as ground truth for future validation purposes. 
    \subsection{Technical Details}
    TableLab is developed in React and Flask while the model (GTE) is developed with TensorFlow. The models preloaded in the demonstration were trained on PubLayNet and PubTabNet, which are large datasets from the scientific papers domain. 
    %    \item Financial Documents (FinTabNet)
    The documents shown for detection and correction labelling are from FinTabNet, which are tables from annual reports of S\&P 500 companies. We demonstrate TableLab's ability to customize models on this new domain with tables that have different styles.  
    
\section{Concrete demo experiences}
    There are three main use cases with our demo. First, users can simply visualize table extraction results with TableLab. Second, AI engineers and scientists can use our tool to quickly create ground truth labels for their documents. Finally, end users can create custom models with their private document collection with our interactive TableLab system.

%\end{itemize}

\bibliographystyle{ACM-Reference-Format}
\bibliography{sample-base}
\end{document}